# Some issues of quantum theory and subquantum processes

V.A. Skrebnev[1] and M.V. Polski[2]

[1] Institute of Physics, Kazan Federal University, 18 Kremlyovskaya str., Kazan 420008, Russia

[2] East-West University, 816 S. Michigan Ave., Chicago, IL 60605, USA

Abstract

This paper critically considers the main interpretations of the wave function and offers an interpretation in which wave function is a consequence of subquantum processes taking place at the level of the organization of matter which underlies the phenomena described by quantum mechanics. We show that this interpretation provides a common ground to explain problematic issues of quantum theory (wave function collapse, electrons' passing through two slits, behavior of electrons within the atom, quantum entanglement). The paper also shows that entropy is determined by the maximum number of macrosystem states which appear as the result of subquantum processes.

## 1. Introduction.

When in 1926 the fundamental Schrödinger equation was created, it generated the problem of offering a convincing interpretation of the meaning of the wave function, which is the solution of this equation. The author of the wave equation himself, Erwin Schrödinger, first thought that we may imagine particles as wave packets and thus abandon the corpuscular view altogether. The flaw of this interpretation quickly became obvious: wave packets inevitably dissipate, which contradicts the obviously corpuscular behavior of particles in electron scattering experiments.

Max Born proposed the probability, or statistical, interpretation [1, 2] which allows to use the Schrödinger equation to describe physical processes without abandoning the corpuscular view. According to Born, every measurement act identifies with a certain probability the value of a physical quantity (e.g. the particle coordinates). The wave function determines this probability. It remains unknown, however, if the wave function contains all needed information about the physical system, because of the inconvenient fact that while the function is the same for every measurement, the result is different every time. Born surmised that the system may possess some properties which are not reflected in the wave function. What these properties are and how they manifest during observation remained unclear.

Copenhagen interpretation, the most popular one in our days, was formulated in 1927. While Born's statistical interpretation does not claim that the probability appears as a result of measurement, Copenhagen interpretation claims that when the system is not being measured, it does not have specific physical properties, and it is only as a consequence of the measurement that the probability of physical values appears. Wave function can predict only the probability of a value appearing as an outcome of measurement. Unlike Born's interpretation, the Copenhagen interpretation argues that the wave function contains all necessary information about the physical system.

Schrödinger found the Copenhagen interpretation unacceptable, because it contradicted his idea of the real quantum-mechanical waves. He was trying to eliminate quantum leaps and other elements of discreteness. He continued to insist on the wave nature of the electron and to treat the electron within the atom as a negatively charged cloud. Schrödinger finally joined the probabilistic interpretation of the nature of waves in 1950. In his paper "What is an elementary particle?" [3] he wrote: "We now say that all waves, including light, are better viewed as 'probability waves'. They are just a mathematical construct to calculate the probability of finding the particle". Schrödinger never accepted the Copenhagen interpretation and till the end of his life he insisted that the wave function should have a real physical meaning.

Many physicists have objected to Copenhagen interpretation, because it treats the wave function as just an auxiliary mathematical tool (and not a physical reality), whose only purpose is to allow us to calculate probability. Einstein exclaimed in a conversation with Abraham Pais: "Do you really think the moon isn't there if you aren't looking at it?" [4].

We will not discuss other interpretations of wave function here (like many worlds interpretation, Bohm interpretation, etc). The fact that so many exist testifies to the absence of a satisfactory one. This is why many physicists lean towards the instrumentalist interpretation, best summed up in the succinct slogan "Shut up and calculate!" [5].

Like many physicists, we do not agree with any of the commonly known interpretations. In [6] the authors introduce the notion of the processes which provide for the formation of wave function. We accept that the physical system exists and lives its life regardless of our observations of it. The system may be in stationary external conditions and be described by a certain wave function, but the coordinates of the particles comprising the system may change. In our view, for a particle to be found in a point in space, it must actually be there at the moment when it is found. The system energy may acquire the values equal to the eigenvalues of the energy operator when the quantum-mechanical average energy value remains the same. In order for the measurements to show a certain energy value, this must be the value of the system energy at the moment of the measurement. Obviously, every value of the system's physical characteristics (e.g. the particle coordinates or energy) corresponds to a certain physical state of the system. When the values of the system's physical characteristics change, it means the system's physical states change. A change in physical states is possible only as a result of physical processes. The wave function which describes the probabilities of the values of various physical characteristics reflects the results of such processes. The processes which form the wave function must exist because there is no other way for the wave function to form. However the processes themselves, naturally, cannot be described by quantum mechanics or by the modern quantum field theory, because these processes take place at the level of the organization of matter which underlies the phenomena described by quantum mechanics, i.e. in the so-called vacuum.

We proceed from a notion of vacuum which is broader than the notion of physical vacuum usually used in the quantum field theory. According to this notion, vacuum is the material medium which generates all matter in the world and sets the parameters for the matter with which our world is built. Accordingly, all parameters of physical systems which are described by quantum mechanics are determined by the vacuum. We will use the term "subquantum processes" for the processes within the vacuum which form the wave function.

The subquantum processes generate system states corresponding to various values of physical quantities. Wave function determines the distribution of the probability of these states. This probability does not appear as a result of measurement, but exists independently of measurement.

To be able to speak of a wave function on the usual time scale, subquantum processes must be extremely fast. This means that to describe them we need a time scale whose graduation is many orders smaller than the usual [6]. Let us call it the subquantum scale. Wave function is the result of averaging over a time period which is extremely short on the usual scale and rather long on the subquantum scale. This interpretation allows to consider wave function as a real physical phenomenon and not as an auxiliary mathematical construct for calculating probability. In our paper we show that the concept of subquantum processes provides a common ground to explain problematic issues of quantum theory.

In section 2 we apply the concept of subquantum processes to discuss the wave function collapse, to consider the electrons' passing through two holes, to clarify the behavior of electrons in the atom and to interpret quantum entanglement.

In section 3 we show that entropy is determined by the maximum number of system states which result from subquantum processes.

We hope that the reader will see that subquantum processes are a productive approach to understanding the fundamentals, problems and paradoxes of quantum theory.

## 2. Wave function collapse, electrons' passing through two holes, behavior of electrons in atoms, entanglement

*2.1 The collapse of wave function during observation*

2.1.1. As a result of subquantum processes, a particle manifests in a certain point in space; if at that moment it happens to be "captured" by a measurement tool, then this particle will not appear in other points in space with the probability described by its wave function before the "capture". This means that the wave function has collapsed.

The movement of the electron in free space is described by the wave equation. In the context of subquantum processes it means that the vacuum makes the electrons manifest in points of space with the probability determined by the square module of the wave function in those points. When an electron at some point of space encounters an obstacle capable of interacting with it, the electron may be absorbed by obstacle, and cause measurable changes in the obstacle. Experience shows that the vacuum will not generate a replacement electron instead of the one that was captured. Naturally, in this case the electron wave, together with the wave function which describes it, will cease to exist. The wave function of the electron will collapse.

Unfortunately, we do not know of methods which can find a particle in a point in space without causing the collapse of the wave function. However, such methods may exist or may be developed.

2.1.2. In compliance with quantum mechanics the function of the system state can be represented as:

$$\psi = \sum_n c_n(t)\psi_n, \tag{1}$$

where $\psi_n$ are eigenfunctions of system Hamiltonian $H$.

Quantum mechanical average system energy, i.e. its total energy, equals

$$E = \langle\psi|H|\psi\rangle = \sum_n |c_n(t)|^2 E_n. \tag{2}$$

Normalization requirement gives us the following:

$$\sum_n |c_n|^2 = 1. \tag{3}$$

According to quantum mechanics, the value $|c_n(t)|^2$ in the equation (2) determines the probability of system energy being equal to $E_n$. However, the value $|c_n(t)|^2$ cannot be considered the probability of eigenstates $\psi_n$ (see textbooks on quantum mechanics). Truly, according to (1) the quantum mechanical average value of any observable $\hat{A}$ is

$$\bar{A} = \langle\psi|\hat{A}|\psi\rangle = \sum_{n,m} c_n^* c_m A_{nm} e^{-i(E_m - E_n)\frac{t}{\hbar}}, \tag{4}$$

$$A_{nm} = \langle\psi_n|\hat{A}|\psi_m\rangle$$

If the value $|c_n(t)|^2$ determined the probability of eigenstates $\psi_n$, then we would have had:

$$\bar{A} = \sum_n |c_n(t)|^2 A_{nn}. \tag{5}$$

It follows that if during measurement we find the system, with a certain probability, in a state with energy $E_n$, it does not mean that we find it with the same probability in an eigenstate corresponding to that energy.

At the same time it is obvious that some physical state of the system must correspond to every different $E_n$. We posit that these states appear as a result of subquantum processes. Let us emphasize that these states should not be confused with the system's eigenstates which are described by wave functions – the eigenfunctions of the system's Hamiltonian (it may be supposed that the combination of all states with this particular energy forms the eigenfunction).

Experience shows that the measurement tool causes disturbances when it locates the system in one of the states with energy $E_n$ caused by the subquantum processes, and thus the system can no longer be described by the wave function which existed before the measurement, which means the collapse of the wave function.

We think when creating the mathematical models of the wave function collapse it is useful to have in mind the physical processes which form the wave function and are responsible for its collapse.

*2.2. Electrons' passing through two holes*

The textbook "The Feynman Lectures on Physics" [7] begins the description of quantum behavior of micro particles by analyzing the electrons' passing through two holes.

If we use the classical ideas about a particle moving in space, and if we assume that particles go through either hole 1 or hole 2, then the probability $P_{12}$ of finding the electrons in some point after they pass the holes should be equal to the sum of probabilities $P_1$ and $P_2$ of finding them after passing through each of the holes, i.e. $P_{12} = P_1 + P_2$. However, experiments show that $P_{12} \neq P_1 + P_2$, and distribution of probabilities yields the interferences picture characteristic for the wave processes.

The textbook [7] says: "It is all quite mysterious. And the more you look at it the more mysterious it seems", and later: "When there is nothing in the experiment to disturb the electrons, then one may *not* say that an electron goes either through hole 1 or hole 2". However, because the electrons do go through the holes and are traced by the detector, we are left to think that the electron mysteriously passes through both holes simultaneously.

We think that out interpretation of the wave function can somewhat reduce the level of mysteriousness in the results of these experiments. According to our interpretation, when the electron wave (described by the wave function) passes through two holes, and the interference picture described by the quantum mechanics is observed, it does not mean that the same electron can pass through two holes at the same time. It is known that the electron can be found in one of the holes with the probability determined by the wave function [7]. In our view, this happens because the electron really happens to be in that hole at the moment of the measurement. Our interpretation suggests that the vacuum may, during the time when the electron wave is passing through the holes (which possess definite geometrical dimensions), repeatedly generate an electron in each hole, though not in both holes at the same time. We think it is only in this sense that one may speak about an electron passing through both holes.

*2.3. Electrons within the atom*

If electrons were moving on an orbit within the atom, this would have been a movement with centripetal acceleration. We know from electrodynamics that a charged particle moving with acceleration radiates energy. Accordingly, very soon the electrons would have given off all their energy and would have fallen on the nuclei, and the Universe to which we are so accustomed would have ceased to exist. Fortunately, it does not happen. The electrons do not move around nuclei on orbits. Instead, the vacuum (according to laws which science has not yet discovered) generates electrons in different spots within the atom, with the probability determined by the square module of the wave function. It is meaningless to say which electron disappeared from one place and appeared in another. This means that the electrons in the atom are identical.

Wave functions of a helium atom, which contains two electrons, and the energy corresponding to those functions, can be written as [8, 9]:

$$\Phi_p(r_1.r_2) = \frac{1}{\sqrt{2}}\{\psi_n(r_1)\psi_m(r_2) + \psi_m(r_1)\psi_n(r_2)\}, \ E_p = E_n + E_m + K + A,$$

$$\Phi_o(r_1.r_2) = \frac{1}{\sqrt{2}}\{\psi_n(r_1)\psi_m(r_2) - \psi_m(r_1)\psi_n(r_2)\}, \ E_o = E_n + E_m + K - A,$$

where $\psi_l(r_i)$ are the wave functions of the electrons in the Coulomb field of the nucleus, and $E_n$ and $E_m$ are their energy values in this field,

$$K = e^2 \int \frac{|\psi_n(r_1)|^2|\psi_m(r_2)|^2}{r_{12}} dv_1 dv_2$$

$$A = e^2 \int \frac{\psi_n^*(r_1)\psi_m(r_1)\psi_n(r_2)\psi_m^*(r_2)}{r_{12}} dv_1 dv_2$$

Quantity $K$ has a simple physical meaning: it is the average energy of Coulomb interaction of two electrons in their independent states $\psi_n(r_1)$ and $\psi_m(r_2)$. Quantity $A$ is called the exchange energy and has no analogues in classical mechanics. Exchange energy is the consequence of the particles being identical and of the existence of states $\Phi_p(r_1.r_2)$ and $\Phi_o(r_1.r_2)$ which are the superposition of products of states of two electrons (such states are usually called "the entangled states" in modern physics).

The states $\Phi_p(r_1.r_2)$ and $\Phi_o(r_1.r_2)$ are the states of parahelium and orthohelium. If the probability of the value of the electron coordinates appeared only as a consequence of observation, then the existence of parahelium and orthohelium would have been only the consequence of observation. However, parahelium and orthohelium, like the Moon, exist in nature independently of observations. Consequently, the probability of the value of the electron coordinates, which allows us to calculate helium atom energy, does not appear as a result of the observations. Instead, it is determined by physical processes, which we call subquantum processes.

## 2.4. Subquantum processes and quantum entanglement

In recent years, the study of entangled states is generating much interest (see, e.g. [10-13]). It is stimulated by the possibility of using quantum entanglement in computation and communication [14].

The wave function of an entangled system is the sum of products of the wave functions of the parts of the system. For example, the wave function of an entangled state with zero spin of the pair of particles possessing spin is:

$$\Phi = \frac{1}{\sqrt{2}}(\psi_1^+ \psi_2^- - \psi_1^- \psi_2^+) \qquad (6)$$

Here $\psi_i^+$ and $\psi_i^-$ (i = 1, 2) are the states where a particle has a clockwise and a counterclockwise spin on a certain axis.

The entangled particles are "one organism" and cannot be considered separately. The measurement results of the state of the entangled particles correlate a priori. For instance, if the total spin of the entangled particles is zero, then when one particle is found to have a clockwise spin on a certain axis, the other particle will be found to have a counterclockwise spin on the same axis. It appears as if the particles "know" about each other's state. Most interestingly, this "knowledge" is preserved even when the particles are separated by such a distance that information about the particle states would have to travel from one particle to another much faster than the speed of light in order to reach it between the measurements of the states of each particle. This is why Einstein was not happy with the concept of entanglement and called such transfer of information "spooky action at a distance" [15].

Because a pair of entangled particles represents "one organism", it is natural to infer that the vacuum simultaneously generates pairs of particles whose wave functions are entangled. If the total spin of a pair of entangled particles is zero, then finding one particle in a clockwise spin state will automatically mean finding the other one in a counterclockwise spin state. In this case it is meaningless to talk about transfer of information from one particle to another, i.e. about the speed of Einstein's spooky action at a distance. Attempts to measure this speed should result in enormous values. For instance, in experiments [16,17] lower boundary of speed of Einstein's spooky action in entangled photon pairs was 4 orders of magnitude of the speed of light (note that these experiments did not measure movement of a mass above the speed of light, which, of course, would have contradicted the theory of relativity).

If the measurement process does not change the states of entangled particles beyond recognition, then the results for different particles will be correlated. This will violate the Bell's inequality [18]; this violation was demonstrated in a significant number of experiments (see, e.g. [12, 16, 17]). Experience shows that the vacuum preserves the "one organism" of entangled particles even at great distances (how the vacuum does it, neither Schrödinger nor Einstein would have told us). A system consisting of two particles whose state is described by an entangled wave function is not a simple "sum" of those particles.

The wave function (which appears as a result of subquantum processes in the vacuum) is a non-local quantity and the distance between the particles is irrelevant.

### 3. Subquantum processes and the entropy

In [6] we used the method of the most probable distribution [19, 20] to derive canonical distribution as the most probable distribution of macrosystem states which result from subquantum processes. In [6] those processes are called hidden internal processes, the system states with energy $E_n$ which appear as their result were called

"instantaneous" states ($E_{ni}$-states), and the actual appearance of these states in the macrosystem was called "visits of $E_{ni}$-states".

Following [6], let us use $N$ for the number of the system's cumulative visits of its "instantaneous" energy states over time $t$ and let $v_n$ be the number of visits of $E_{ni}$-states, corresponding to the energy $E_n$, over this time. Obviously,

$$N = \sum_n v_n \tag{7}$$

Let's introduce the value

$$E_t = \sum_n v_n E_n. \tag{8}$$

Numerous "configurations" determined by various sets of numbers of visits $v_n$ correspond to the value $E_t$. Each "configuration" may be realized in $P$ ways corresponding to the number of permutations of the visits:

$$P = \frac{N!}{v_1! v_2! \ldots v_l! \ldots}. \tag{9}$$

To find the maximum of the function $P$, [6] used the Lagrange method, which involves finding the extremum of the function

$$\ln P - \alpha \sum_i v_i - \beta \sum_i v_i E_i. \tag{10}$$

when conditions (7) and (8) are observed. In (10) $\alpha$ and $\beta$ are the Lagrange multipliers.

Taking into account the Stirling approximation

$$\ln(m!) = m(\ln m - 1), \tag{11}$$

we find that the maximum of function $P$ corresponds to the most probable distribution

$$v_n = \frac{N e^{-\beta E_n}}{\sum_n e^{-\beta E_n}}. \tag{12}$$

and canonical distribution

$$\rho(E_n) = \frac{v_n}{N} = \frac{e^{-\beta E_n}}{\sum_n e^{-\beta E_n}}. \tag{13}$$

It is only possible to use formula (13) to calculate the average values of a macrosystem's physical characteristics if the states with different values of $E_n$ appear during the observation time with the probability described by this formula. The processes which provide for the appearance of these states (the processes we call subquantum), obviously must happen extremely fast for the standpoint of a timescale familiar to us. As has been noted before, for the subquantum processes we need a time scale whose graduation is many orders smaller than the usual.

Let us show now that entropy is determined by the maximum number of macrosystem states which are generated by subquantum processes.

The total energy of macrosystem

$$E = \sum_n \rho(E_n) E_n \qquad (14)$$

Using equations (9), (11), (12), (13), (14) and taking into account that $\sum_n \nu_n = N$, $\sum_n \rho(E_n) = 1$, for maximum $P$ we receive:

$$lnP_{max} = \ln N! - \sum_n \ln(\nu_n!) = N(\ln N - 1) - \sum_n \nu_n (\ln \nu_n - 1) =$$
$$NlnN - \sum_n \nu_n \ln \nu_n =$$
$$NlnN - \sum_n N\rho(E_n) \ln(N \rho(E_n)) = NlnN -$$
$$N \sum_n (\rho(E_n) \ln N + \rho(E_n) \ln \rho(E_n)) = NlnN -$$
$$NlnN \sum_n \rho(E_n) - N \sum_n \rho(E_n) \ln \rho(E_n) = - N \sum_n \rho(E_n) \ln \rho(E_n) =$$
$$-N \sum_n \frac{e^{-\beta E_n}}{\sum_n e^{-\beta E_n}} \ln \frac{e^{-\beta E_n}}{\sum_n e^{-\beta E_n}} = -N \sum_n \frac{e^{-\beta E_n}}{\sum_n e^{-\beta E_n}} \left( \ln e^{-\beta E_n} - \ln \sum_n e^{-\beta E_n} \right) =$$
$$N \left( \beta \sum_n \rho(E_n) E_n + \sum_n \rho(E_n) \ln \sum_n e^{-\beta E_n} \right) = N \left( \beta E + \ln \sum_n e^{-\beta E_n} \right) =$$
$$-N \ln \frac{e^{-\beta E}}{\sum_n e^{-\beta E_n}} = -N \ln \rho(E),$$

from which it follows:

$$\frac{lnP_{max}}{N} = -\sum_n \rho(E_n) \ln \rho(E_n) = -\ln \rho(E) = S \qquad (15)$$

where $S$ – entropy of system (compare to the formula (61) of Boltzmann's paper [20] and with the formula (7.10) of the textbook [21]).

In the textbook [21] entropy is defined as a logarithm of the number $\Delta \Gamma$ of eigenstates of a macrosystem defined from a condition

$$\rho(E) \Delta \Gamma = 1 \qquad (16)$$

(in [21] $\rho(E)$ it is designated as $w(E)$).

When this condition is valid then

$$ln\Delta\Gamma = -ln\rho(E) = S \qquad (17)$$

It is supposed in [21] that probability of the system to have energy $E_s$ has extremely sharp maximum at $E_s = E$. It is obvious that the change of energy $E_s$ can be caused only by interaction of system with an environment. In the textbook [21] only eigenstates of system are assigned to each value of energy $E_s$. However, the equations (2) and (3) have an enormous number of solutions for $|c_n|^2$ if the number of levels is more than two (see [6]). This means that each value of energy $E_s$ corresponds to numerous system states, and not only eigenstates.

In [21], the probability of eigenstates in the interval $\Delta\Gamma$ is considered the same and equal to $\rho(E)$. Textbook [21] does not discuss the mechanisms of transitions between these states under the influence of environment, nor does it discuss the frequencies of these transitions which would allow us to use some number as the probability of a certain state.

Formula (15) connects entropy to the maximum number of system state realizations as a consequence of subquantum processes. We consider that the states which are described by canonical distribution (13) are the consequence of the macrosystem's tendency towards the maximum freedom in realizing its state with a given total energy, i.e. towards maximum entropy, expressed by (15).

**Conclusion**

In this paper we have proposed to consider that the distribution of the probability of the physical values, which is determined by the wave function, arises from subquantum processes which take place on the level of the organization of matter which underlies the phenomena described by quantum mechanics, i.e. in the vacuum. These processes allow to explain from one position the meaning of the wave function, the collapse of this function during measurement, the electrons' passing through two slits, behavior of electrons in the atom, and the paradoxes of entangled states. We have also shown that entropy is expressed through the number of system states, which are generated by subquantum processes, with the most probable distribution of those states.

It is common knowledge that behind every chance there is a rule. Accordingly, behind the quantum-mechanical probability there must be the rules of the processes in the vacuum. From our point of view, our results call for the need to recognize the existence of the subquantum processes and to study them. The subquantum processes defining behavior of physical systems are one of manifestations of the general properties of a matter in the Universe. The study of these processes will further our knowledge of the laws of the microworld and macroworld.